# Detection of nano scale thin films with polarized neutron reflectometry at the presence of smooth and rough interfaces

Saeed S. Jahromi*, Seyed Farhad Masoudi

Department of Physics, K.N. Toosi University of Technology, 15875-4416, Tehran, Iran
E-mail:
Saeed S. Jahromi, s.jahromi@dena.kntu.ac.ir
Seyed Farhad Masoudi: Masoudi@kntu.ac.ir

By knowing the phase and modules of the reflection coefficient in neutron reflectometry problems, a unique result for the scattering length density (SLD) of a thin film can be determined which will lead to the exact determination of type and thickness of the film. In the past decade, several methods have been worked out to resolve the phase problem such as dwell time method, reference layer method and variation of surroundings, among which the reference method and variation of surroundings by using a magnetic substrate and polarized neutrons is of the most applicability. All of these methods are based on the solution of Schrodinger equation for a discontinuous and step-like potential at each interface. As in real sample there are some smearing and roughness at boundaries, consideration of smoothness and roughness of interfaces would affect the final output result. In this paper, we have investigated the effects of smoothness of interfaces on determination of the phase of reflection as well as the retrieval process of the SLD, by using a smooth varying function (Eckart potential). The effects of roughness of interfaces on the same parameters, have also been investigated by random variation of the interface around it mean position.

# 1  Introduction

Spin-polarized neutron reflectometry is one of the most applicable probes to the study of thin films which have been developed theoretically and experimentally in the past decades. In this technology the reflectivity profile, $R(q)$, of a flat thin film in term of the neutron wave number, $q = 2\pi \sin\theta / \lambda$, where $\lambda$ is the neutron wavelength and $\theta$ is the reflection angle, is measured to obtain the scattering length density of the sample [1].

If only a single reflectivity curve of a thin film is measured, it would not result to a unique solution for the SLD of the sample as more than one SLD can be find which correspond with the same reflectivity curve. In order to solve this problem, the knowledge of the phase angle $\phi$ of the complex reflection amplitude, $r(q)$, would also be decisive in determining the correct structure [2-5].

In order to retrieve the phase information in reflectometry problems, several methods have been developed such as, reference layers [2], and variation of surroundings which were first proposed and tested experimentally by Majkerzak et al. [6].

The reference method is based on three measurements of reflectivity from a composite sample which is consisted of a known and an unknown part. The method at first was presented by using three different known films as reference layers [2], however due to the difficulties of changing the reference layers for a distinct sample the method was later developed by using polarized neutrons and a magnetic film as reference layer [3].

The method of variation of surroundings is based on two measurement of reflectivity from a sample with variable surrounding medium. Such an approach is suited to a substantial number of systems in which the film is in contact with an aqueous reservoir serving as the transmitting medium whose SLD can be readily varied by exchanging heavy for light water. Although the method is practical, it brings about some limitation in variation of the surrounding medium by making us choose just aqueous or gaseous materials [6]. This deficiency then was recovered by using a ferromagnetic substance as substrate and polarized neutrons as incident beam by Leeb et al. [7-8] and then developed and formulated in a straightforward manner by Masoudi et al [9,10].

For a magnetic film, the SLD depth profile is proportional to ($\rho(z) = \rho_n \pm \mu . B$), where $\mu$ is the neutron magnetic moment and $B$ is the magnetic field inside the magnetic layer. The plus and minus signs refer to neutrons parallel and anti parallel to the local quantization. As an example for a magnetic field in the plane of the film, and incident neutrons polarized parallel and anti parallel to the magnetic field, the SLD of the film would increase and decrease by the values $+\mu . B$ and $-\mu . B$ respectively [1]. Based on this knowledge, the surrounding medium of a sample or the reference layers in reference method can readily be varied by using a magnetic material and polarized neutrons [7-10].

In all of the outlined reflectometry methods, it is supposed the interacting potential between neutrons and sample at boundaries to be discontinuous (ideal step-like sample). As we know from a real sample, there are some smearing and roughness at boundaries. Considering the smoothness and roughness of interfaces, would affect the output reflectivity. In this brief report, we have studied the effects of the smoothness and roughness of interfacial potential on reflectivity, polarization of reflected neutrons and phase of the reflection amplitude by using Eckart potential and random variation of boundaries position to consider smoothness and roughness respectively. We have also investigated the stability of the reference method and variation of surroundings method on retrieving the SLD of a thin multilayer system by using Polarized Neutron Reflectometry (PNR). In order to precisely define the smoothness and roughness of optical potential at boundaries, we use the following terminologies: Ideal sample stands for a step-like potential with discontinuity at boundaries (Fig. 1-a), Smoothness stands for the smearing of both interfaces for a layer on a substrate or over another layer (Fig. 1-b) and Roughness stands for the harsh distribution of optical potential of two distinct layer at their interface (Fig. 1-c).

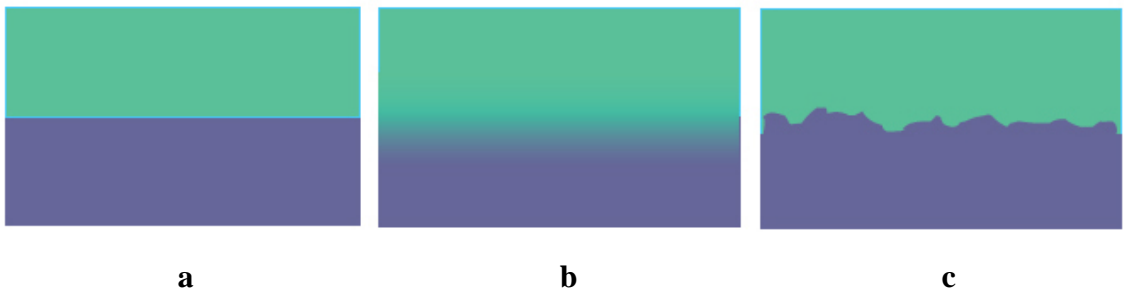

**a**  **b**  **c**

**Figure 1: a**) An ideal step-like sample with discontinuous potential at boundaries. **b**) A non-ideal sample with continuous potential at boundaries (smooth potential at boundary). **c**) A non-ideal sample with roughly distributed potential at boundaries (rough potential at boundary).

The layout of the paper is as follow: in section 2 we will explain the basic principles of reflectometry with polarized neutrons, in section 3 the method of variation of surrounding which was developed by Masoudi et al. is presented and the retrieval of complex reflection coefficient for an unknown free revered film by using a magnetic substrate is discussed. Section 4 covers the method of magnetic reference layers for determining the phase of reflection. In Section 5 we will consider the smoothness of interfacial potential by using Eckart function and the roughness is considered by random variation of boundary position. Finally in section 6 , effects of potential smoothness and roughness on phase of reflection and retrieval of the SLD is investigated by implementing the smooth potential of section 5 into relations of section 3 and 4 by using a numerical simulation for a distinct sample.

## 2  Theory

An alternative exact representation of the complex reflection coefficient, $r(q)$, which is obtained from the solution of one dimensional schrödinger equation, can be derived by using the transfer matrix method. The elements of the 2×2 transfer matrix, $A(q), B(q), C(q), D(q)$ for a thin film which is mounted on top of a substrate is expressed as[2-4]:

$$\begin{pmatrix} 1 \\ ih \end{pmatrix} t e^{ihqL} = \prod_{j=N}^{1} M_j \begin{pmatrix} 1+r \\ if(1-r) \end{pmatrix},$$

$$M_j = \begin{pmatrix} A & B \\ C & D \end{pmatrix} = \begin{pmatrix} \cos q_j d_j & \dfrac{q_0 \sin q_j d_j}{q_j} \\ \dfrac{-q_j \sin q_j d_j}{q_0} & \cos q_j d_j \end{pmatrix}$$

(1)

where $r$ and $t$ are the reflection and transmission coefficients, $f$ and $h$ are the refractive index of the fronting and backing medium respectively, which can generally be written as $n = (1 - 4\pi \rho_n / q_0^2)^{1/2}$ for $n=f$ and $h$. In the elements of matrix $M_j$, $q_j = (q_0^2 - 4\pi \rho_j)^{1/2}$ is the neutron wave number in the $j$ th layer, $q_0$ is the neutron wave number in vacuum and $d_j$ is the thickness of $j$ th layer. Solving eq(1) for reflection coefficient, we have:

$$r(q) = \frac{\beta^{fh} - \alpha^{fh} - 2i\gamma^{fh}}{\beta^{fh} + \alpha^{fh} + 2} \tag{2}$$

where

$$\begin{aligned} \alpha^{fh} &= hf^{-1} A^2 + (fh)^{-1} C^2 \\ \beta^{fh} &= fh B^2 + fh^{-1} D^2 \\ \gamma^{fh} &= hAB + h^{-1} CD \end{aligned} \tag{3}$$

where the superscript $fh$ denotes a sample with fronting and backing medium of refractive indexes $f$ and $h$ respectively. The reflectivity, $R(q) = |r(q)|^2$, can be related to the elements of the transfer matrix in term of new quantity $\sum(q)$, as [2]:

$$\sum(q) = 2 \frac{1+R}{1-R} = \alpha^{fh} + \beta^{fh} \tag{4}$$

The components of polarization of incident and reflected neutrons are related to the reflection amplitude of a sample, as follow [7,8]:

$$\begin{aligned} P_x + iP_y &= \frac{2 r_+^* r_- (P_x^0 + i P_y^0)}{R_+(1+P_z^0) + R_-(1-P_z^0)} \\ P_z &= \frac{R_+(1+P_z^0) - R_-(1-P_z^0)}{R_+(1+P_z^0) + R_-(1-P_z^0)} \end{aligned} \tag{5}$$

where $P_i^0$, $i=x,y$ and $z$ are the components of polarization of incident neutrons, $r_\pm$ are the reflection coefficient of the sample for incident neutrons parallel (plus sign) and anti parallel (minus sign) to the magnetization of the magnetic layers and $r_+^*$ is the complex conjugate of $r_+$.

## 3   Determining the phase by variation of surroundings

For an unknown sample which is mounted on top of a magnetic substrate, Eq. (4) can be written as [10]:

$$\sum_\pm = f h_\pm^{-1} \tilde{\alpha}_\pm^{ff} + h_\pm f^{-1} \tilde{\beta}_\pm^{ff} \tag{6}$$

where $h_\pm$ is the refractive index of the magnetic substrate, (+) and (−) signs correspond to the incident neutrons parallel and anti parallel to the local quantization, the tiled denotes the mirror-reversed unknown film which is the interchange of the diagonal elements of the corresponding transfer matrix $(A \leftrightarrow D)$, and the superscript $ff$ represents vacuum surrounding at both sides.

Using equations (2), (5) and (6), which relate the polarization of incident and reflected neutrons to $\sum_\pm$, we can extract three unknown parameters, $\tilde{\alpha}_\pm^{ff}$, $\tilde{\beta}_\pm^{ff}$ and $\tilde{\gamma}_\pm^{ff}$ which completely determine the complex reflection amplitude and phase of the free reversed unknown film [10]. The parameter $\tilde{\gamma}_\pm^{ff}$ is expressed as:

$$\tilde{\gamma}_\pm^{ff} = \left(\frac{h_+ - h_-}{h_+ + h_-}\right) \zeta_p \tag{7}$$

where

$$\zeta_p = \frac{(P_x P_y^0 - P_y P_x^0)(1 - P_z^{0^2})}{P_x P_x^0 - P_y P_y^0 - (P_x^{0^2} + P_y^{0^2})(1 - P_z P_z^0)} \tag{8}$$

By the knowledge of $\sum_\pm$, the two other unknown parameters is derived from equation (6):

$$\tilde{\alpha}_\pm^{ff} = \frac{h_+ h_- (h_+ \sum_- - h_- \sum_+)}{f(h_+^2 + h_-^2)} \tag{9}$$

$$\tilde{\beta}_\pm^{ff} = \frac{f(h_+ \sum_+ - h_- \sum_-)}{(h_+^2 + h_-^2)} \tag{10}$$

The quantities $\sum_\pm$, can be determined by the solution of the following quadratic equation [10]:

$$\sum_\pm^2 \pm \frac{(h_+ - h_-)^2}{h_+ h_-} \zeta_p' \sum_\pm = 4 + 2\frac{(h_+ - h_-)^2}{h_+ h_-} \zeta'\left(\frac{P_z P_z^0 - 1}{P_z - P_z^0}\right) \tag{11}$$

where

$$\zeta_p' = \frac{(P_x^{0^2} + P_y^{0^2})(P_z - P_z^0)}{(P_x P_x^0 - P_y P_y^0)(1 - P_z^{0^2}) + (P_x^{0^2} + P_y^{0^2})(P_z P_z^0 - 1)} \tag{12}$$

Eq. (11) has two different solutions. The physical solution is selected from the fact that $\sum_{\pm} \geq 2$ [9,10].

The method provides us with several procedures in determining the reflection coefficient by choosing to measure three possible selection of $R_+, R_-, P_x, P_y$ and $P_z$ [10]. After extracting the unknown parameters, $\tilde{\alpha}_{\pm}^{ff}$, $\tilde{\beta}_{\pm}^{ff}$ and $\tilde{\gamma}_{\pm}^{ff}$, the reflection coefficient of the free reversed unknown film, $\tilde{r}_{\pm}^{ff}$, is determined and phase of the reflection is expressed as:

$$\varphi = Arc\tan(\frac{-2\tilde{\gamma}_{\pm}^{ff}}{\tilde{\beta}_{\pm}^{ff} - \tilde{\alpha}_{\pm}^{ff}}) \tag{13}$$

The missing data below the critical value of $q_c$, where $q_c = 4\pi\rho_{h_+}$, can be determined by the fact that $\tilde{r}_{\pm}^{ff} \rightarrow -1$ as $q \rightarrow 0$ and $\tilde{r}_{\pm}^{ff}(-q) = \tilde{r}_{\pm}^{*ff}(q)$ [10,12]. Thus, $\tilde{r}_{\pm}^{ff}$ is known over the whole range of $q$.

## 4  Determining the phase by reference method

As we mentioned previously on section 1, the method of reference layers is based on the measurement of reflectivity from a composite sample which one of its parts is known and the other is unknown. Based on this definition, the transfer matrix of the sample can be written as multiplication of transfer matrix of the known and unknown part. The $\sum_{\pm}$ parameters for the reference method can be written as [11]:

$$\sum_{\pm}(q) = \beta_{k\pm}^{fh}\tilde{\alpha}_u^{ff} + \alpha_{k\pm}^{fh}\tilde{\beta}_u^{ff} + 2\gamma_{k\pm}^{fh}\tilde{\gamma}_u^{ff} \tag{14}$$

where the subscripts "$k$" and "$u$" refers to known and unknown part of the sample and the tilde represent the mirror reversed potentials. After some straightforward calculation, using Eq. (2), (5), (14) we have [11]:

$$\frac{P_x P_x^0 + P_y P_y^0}{P_x^{0^2} + P_y^{0^2}} = 1 + 2\frac{\zeta_k - P_z^0(\Sigma_+ - \Sigma_-)}{\Sigma_+\Sigma_- + 2P_z^0(\Sigma_+ - \Sigma_-) - 4} \tag{15}$$

$$\frac{P_x P_y^0 - P_y P_x^0}{P_x^{0^2} + P_y^{0^2}} = 1 + 2\frac{2(c_{\beta\gamma}\tilde{\alpha}_u^{ff} + c_{\gamma\alpha}\tilde{\beta}_u^{ff} + c_{\beta\alpha}\tilde{\gamma}_u^{ff})}{\Sigma_+\Sigma_- + 2P_z^0(\Sigma_+ - \Sigma_-) - 4} \tag{16}$$

$$P_z - P_z^0 = 1 + 2\frac{2(1 - P_z^{0^2})(\Sigma_+ - \Sigma_-)}{\Sigma_+\Sigma_- + 2P_z^0(\Sigma_+ - \Sigma_-) - 4} \tag{17}$$

where

$$\zeta_k = 2(1 + \gamma_{k+}^{fh}\gamma_{k-}^{fh}) - (\alpha_{k+}^{fh}\beta_{k-}^{fh} + \beta_{k+}^{fh}\alpha_{k-}^{fh}) \tag{18}$$

and

$$c_{ij} = i^{fh}_{k+} j^{fh}_{k-} - j^{fh}_{k+} i^{fh}_{k-} \qquad (19)$$

for $i$ and $j$ = '$\alpha$', '$\beta$' and '$\gamma$'. The parameters $c_{ij}$ and $\zeta_k$ are known from the elements of the transfer matrix of the known reference layer.

The numerator on the right side of Eq. (16) depends on the unknown parameters. Furthermore, the dependence of Eqs. (15) and (17) to the unknown parameters is encoded in the parameters $\Sigma_+$ and $\Sigma_-$. Using this dependency and the polarization of incident and reflected neutrons we have:

$$\Sigma_\pm^2 \mp \frac{\zeta_k}{\zeta_p} \Sigma_\pm - (4 + 2\frac{\zeta_k(1-P_z P_z^0)}{\zeta_p(P_z - P_z^0)}) = 0 \qquad (20)$$

where

$$\zeta_p = P_z^0 + \frac{1-P_z^{0^2}}{P_z - P_z^0}(\frac{P_x P_x^0 - P_y P_y^0}{P_x^{0^2} + P_y^{0^2}} - 1) \qquad (21)$$

Similar to Eq. (11) in section 3, the quadratic equation of (20) also has two different solutions which only one of them satisfies the physical condition [9,10]. By knowing $\Sigma_+$, $\Sigma_-$ and using Eq. (16), we have three set of linear equations which depend on unknown parameters. By solving these set of equations by using matrix algebra we have:

$$\begin{pmatrix} \tilde{\alpha}_u^{ff} \\ \tilde{\beta}_u^{ff} \\ \tilde{\gamma}_u^{ff} \end{pmatrix} = T^{-1} \begin{pmatrix} \Sigma_+ \\ \Sigma_- \\ \dfrac{\zeta_k(P_x P_y^0 - P_y P_x^0)(1-P_z^{0^2})}{\zeta_p(P_z - P_z^0)(P_x^{0^2} - P_y^{0^2})} \end{pmatrix}, \qquad T = \begin{pmatrix} \beta_{k+}^{fh} & \alpha_{k+}^{fh} & 2\gamma_{k+}^{fh} \\ \beta_{k-}^{fh} & \alpha_{k-}^{fh} & 2\gamma_{k-}^{fh} \\ c_{\gamma\beta} & c_{\alpha\gamma} & c_{\alpha\beta} \end{pmatrix} \qquad (22)$$

The parameters $\tilde{\alpha}_u^{ff}, \tilde{\beta}_u^{ff}$ and $\tilde{\gamma}_u^{ff}$ denote the free reversed unknown sample

## 5  Smooth and rough interfacial potentials

As we mentioned in chapter 1, there is some smearing at boundaries in real samples and the SLD varies smoothly and continuously from one layer to another. In this paper we have investigated the smoothness of interfacial potential by choosing Eckart potential which denotes the smooth variation of SLD at boundaries [13]. The roughness of boundaries has also been considered by random variation of the interface position [8].

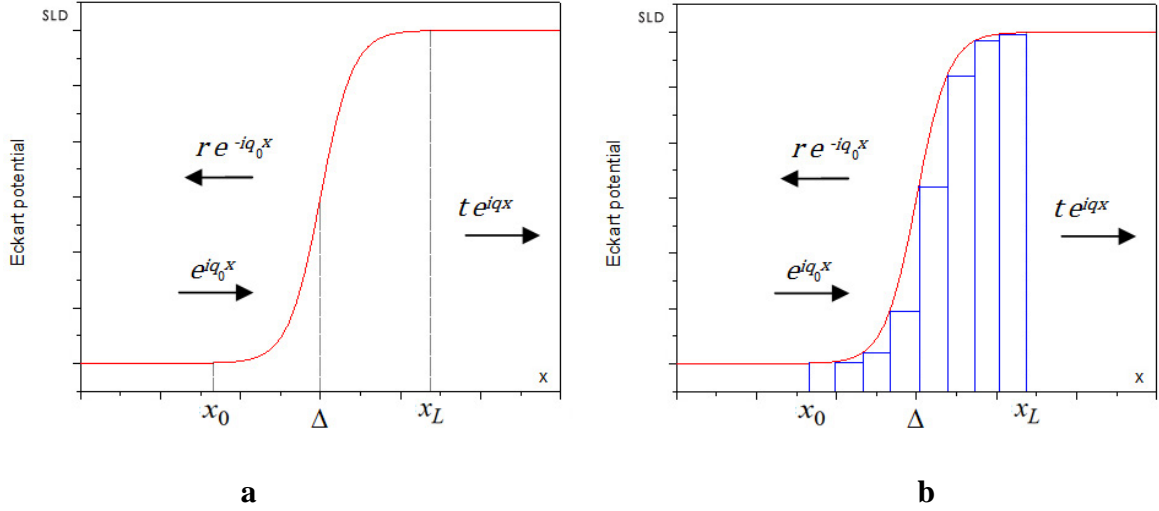

**Figure 2: a)** Smooth varying Eckart potential. **b)** The smooth area is divided into several small rectangular layers. The total transfer matrix of the smooth area is determined by multiplication of the transfer matrix of rectangular layers.

The smooth variation of the potential at boundaries can be expressed by the so-called Eckart potential which is one of the most applicable functions in nuclear physics specially for considering the smearing of nuclear radius. The variation of the SLD corresponding to the Eckart potential is shown as follow [13]:

$$\rho(x) = \rho_1(x) + (\rho_1(x) - \rho_2(x)) / [1 + \exp(\frac{x-\Delta}{b})] \qquad (23)$$

where $b$ is the smoothness factor. In other word, $b$ is thickness of the smoothly varying area across the interface. $\Delta$ is also the mean position of the interface or the turning point of the Eckart potential (Fig. 2-a). As illustrated in Fig. 2-b, the transfer matrix of the smooth part ($X_0$ to $X_L$), can be determined, using step method, in which, we divide the smooth part into several small rectangular layers and multiply the transfer matrix of individual layer to gain the total transfer matrix.

In order to investigate the roughness of interfaces, we randomly change the position of boundaries around its mean position, by adding (subtracting) a certain value to (from) the mean position of the interface. This certain value is determined by randomly generated numbers under a normal distribution. In order to produce random numbers with Gaussian distribution, we use the well known Box Muller method [19] in which, two set of random numbers with standard distribution ($y_1$, $y_2$) are used as input in Eqs. (24), (25) and two new sets of numbers with normal distribution ($x_1$, $x_2$) are generated as follow:

$$x_1 = \sqrt{-2\delta^2 \ln(y_1)} \cos(2\pi y_2) \qquad (24)$$

$$x_2 = \sqrt{-2\delta^2 \ln(y_1)} \sin(2\pi y_2) \tag{25}$$

where $\delta$ is the half-with of the Gaussian distribution. Each of the sets of $x_1$ or $x_2$ can be added to the mean position of the boundary and randomly changes the boundary's position. This approach to the roughness of boundaries truthfully makes sense in one dimension. The method is the same as the one which is used by Leeb in reference [8].

## 6 Numerical Example

To investigate the effects of the smoothness and roughness of the interfacial potential on the output results of the variation of surroundings and reference method, we consider a bilayer sample composed of 30 nm thick Au over a 20 nm thick Cr with the SLD of 4.46 and 3.03 $\times 10^{-4}$ nm$^{-2}$ for Au and Cr respectively. As shown in Fig. 3, in the method of variation of surroundings, it is supposed the sample to be over a Cobalt substrate with the SLD of 6.44 and -1.98 $\times 10^{-4}$ nm$^{-2}$ for plus and minus magnetization, respectively. We have also chosen a 20 nm Cobalt magnetic film as reference layer in reference method which is mounted between the sample and a silicon substrate (Fig. 4). The polarization of incident neutrons is as follow: $P_z^0 = 0.8$ and $P_y^0 = 0.2$. It is supposed the magnetic field of the Cobalt films to be along the $z$ axis in both reference method and variation of surroundings.

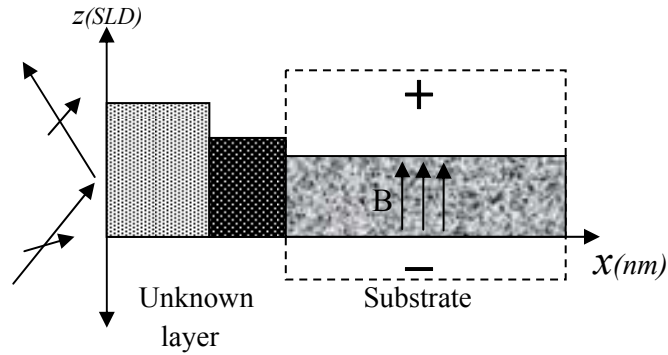

**Figure 3:** Arrangement of a sample for investigating the smoothness and roughness of interfacial potential in the method of variation of surroundings. Dashed line represents the effective potential experienced by neutrons parallel and anti parallel to the magnetic field **B** inside the substrate.

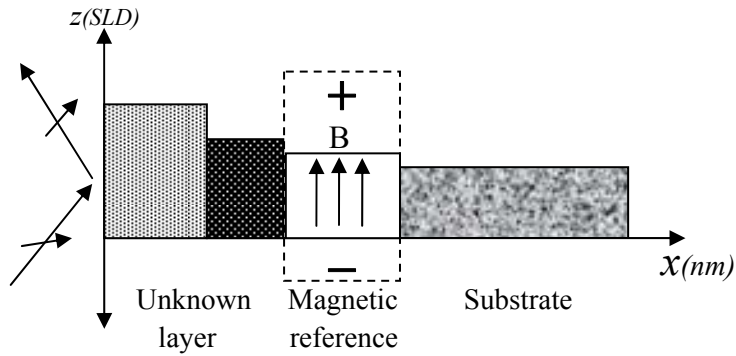

**Figure 4:** Arrangement of a sample for investigating the smoothness and roughness of interfacial potential in the reference method. Dashed line represents the effective potential experienced by neutrons parallel and anti parallel to the magnetic field **B** inside the magnetic reference layer.

Fig. 5-a, illustrates the interacting optical potential of fig. 3 after implementation of Eckart function with interface width of $b=5A^o$ at interfaces. The dotted lines locate the mean position of the interface and continuous variation of optical potential at boundaries is clearly demonstrated. As we mentioned in the previous chapter, the roughness of boundaries are considered by random variation of the interface position around its mean position. The distribution function of interface position for one of the interfaces (interface of Au and Cr) is plotted in fig. 5-b. As it is shown in the figure, the interface variation under a normal distribution with approximately 0.5 nm variation around its mean position (30 nm), is obvious.

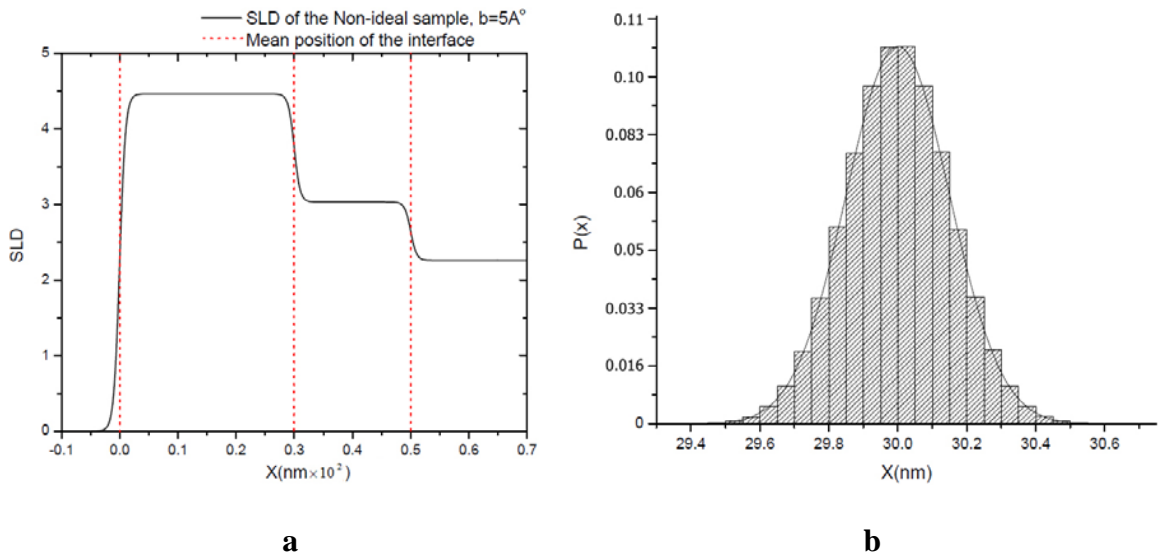

a          b

**Figure 5:** **a**) The interacting optical potential of sample of fig. 3 with consideration of smoothness at boundaries. The interface width is: $b=5A^o$. **b**) Random variation of mean position of the interface at 30nm under a normal distribution.

As it is shown in Fig. 6-a, the polarization curve of reflected neutrons for the Eckart potential with smoothness factor $b=5A^o$, is illustrated. Fig. 6-b, demonstrates the reflectivity curve and finally Fig 6-c, depicts the phase curve at the presence of smoothness, using the variation of surroundings method. In addition, Fig, 7-a and b, demonstrate the output results of reference method for the reflectivity and phase of reflection, with the smoothness factor, $b=5A^o$, respectively.

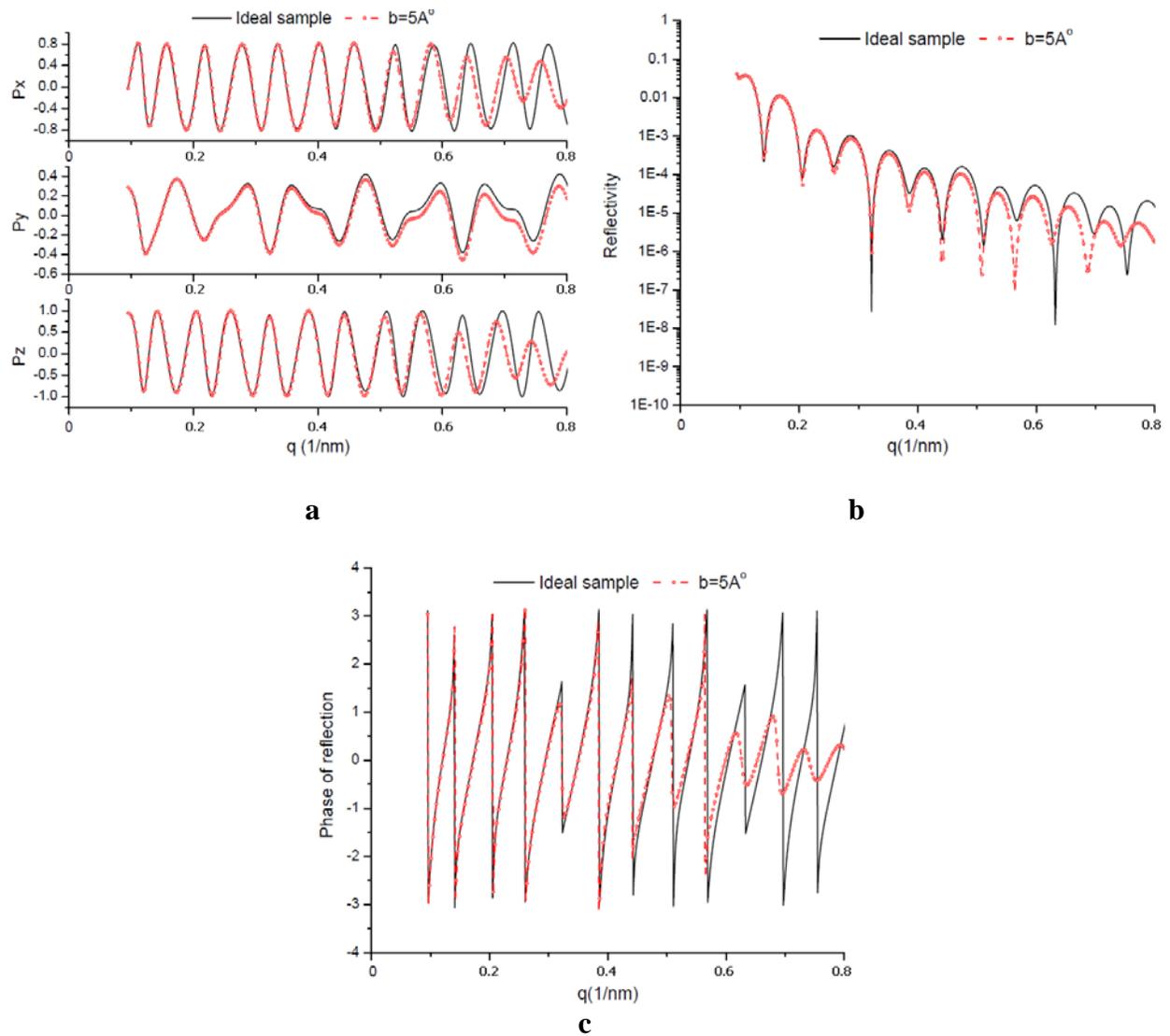

**Figure 6:** Output results of variation of surroundings method at the presence of smoothness **a)** Polarization of the reflected neutrons, **b)** reflectivity curve and **c)** phase of reflection, for the Eckart potential with smoothness factor; $b=5A^o$.

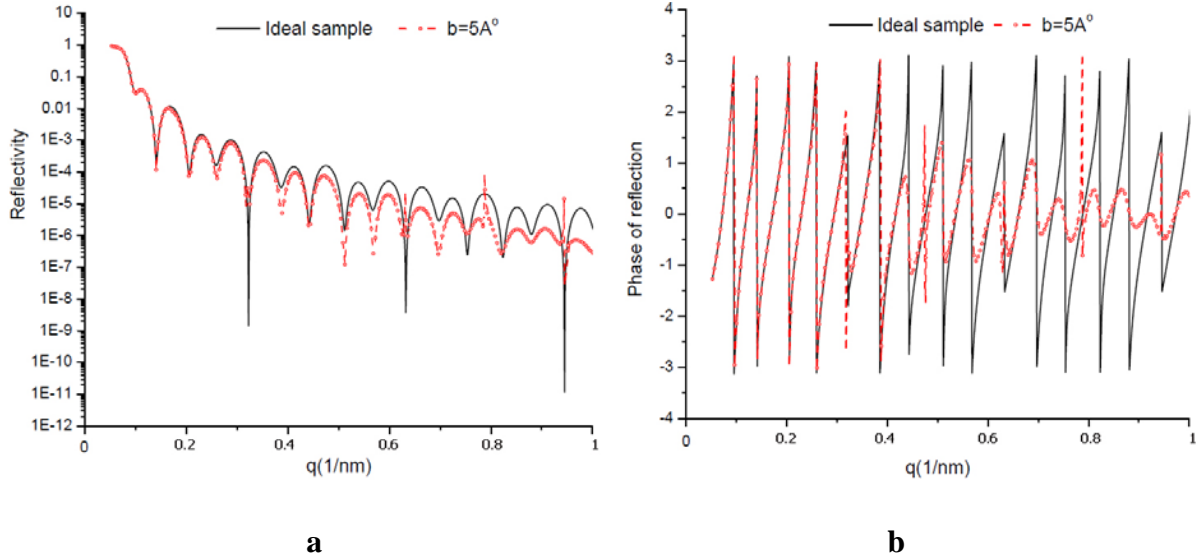

**Figure 7:** Output results of reference method at the presence of smoothness **a)** reflectivity and **b)** phase of reflection, for the Eckart potential with smoothness factor; $b=5A^o$. The abrupt noises are clear in both reflectivity and phase curves, particularly at large wave numbers ($q>0.6$)

As it is illustrated in Figure 6 and 7, the circled curves show the output data at the presence of smoothness. The effects of smoothness on output results is clear in all of the figures particularly at large wave numbers, while the data for small wave numbers truly correspond with none-smooth data. Fig 7-a and b, however, demonstrate some abrupt noises in both the reflectivity and phase curves of reference method. These noises are due to the abrupt change in some of the elements of the $T$ matrix, Eq. (22), in reference method. The intensity and distribution of these noises along the different range of wave numbers are extremely relevant to the smoothness factor, thickness of the reference layers and the range of neutron wave numbers. In order to reduce or completely remove the noises of the reflectivity and phase curve, we have proposed a method in [18], based on which the noises can be eliminated or shifted to the large range of wave numbers by choosing a proper thickness for the reference layers. Other probable remaining noises can be removed by extrapolation or ignoring the data of large wave numbers ($q>0.6$) which are not compulsory in retrieving the SLD of the sample [17,18]. Fortunately the method of variation of surroundings with polarized neutrons is totally free of any noise and absolutely stable at the presence of smoothness which makes the retrieval process of SLD, more fast and reliable.

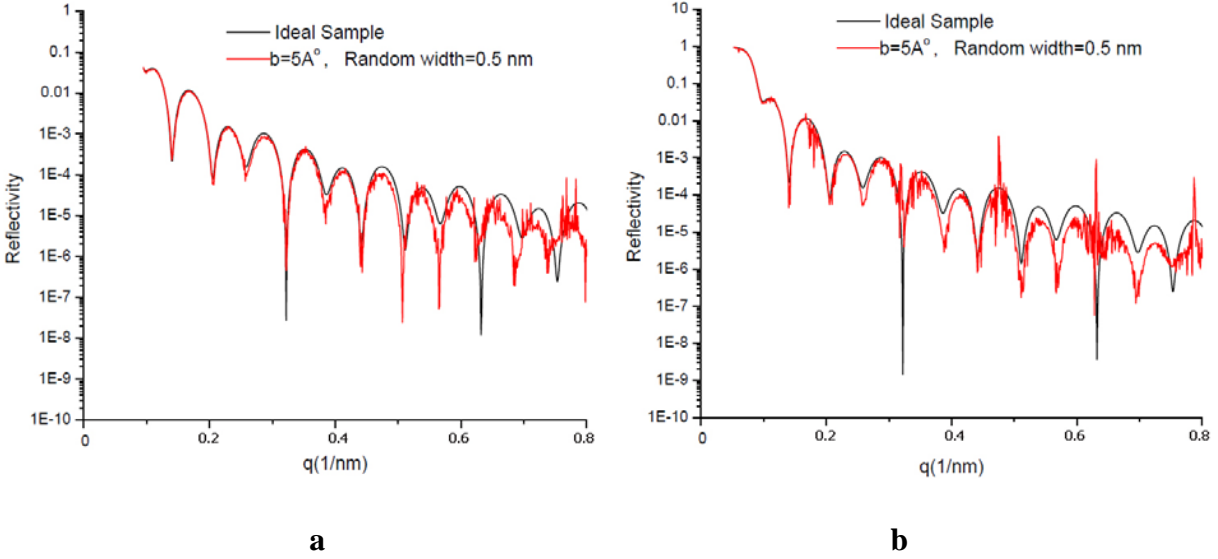

**Figure 8:** Reflectivity curve of the samples of Fig. 3 and 4 by considering smoothness ($b=5A^o$) and roughness, using **a**) variation of surroundings and **b**) reference method.

In order to investigate the effects of roughness of interfaces, we have randomly changed the position of the interfaces at the same time with consideration of smoothness with smoothness factor $b=5A^o$.

According to what was explained in chapter 5 for implementation of roughness, we use two sets of random numbers which are generated by RAN function in Fortran90. The generated numbers by this function have standard distribution. Using the numbers as input in Eq. (24), (25) and also considering the half-with $\delta=0.02$, the new set of random numbers with normal distribution will be generated.

Fig 8-a, b, shows the reflectivity curves of the sample at the presence of smoothness and roughness for variation of surroundings and reference method, respectively. As it is illustrated in both figures, the noises are existed, specifically at large wave numbers. However in Fig. 8-a (variation of surroundings), noises have appeared at large wave numbers ($q>0.6$), while for the reference method (Fig. 8-b), distribution of the noises is over the whole range of wave numbers even at $q<0.6$, which makes the data useless to retrieve the SLD of the sample.

In order to retrieve the SLD of an unknown sample, the data of real and imaginary parts of the reflection coefficient in whole range of wave numbers, particularly below the $q_c$, are needed. By using these data as input for some useful codes like the one which is developed by P. Sacks [16] based on Gel'fan-d-Levitan integral equation [14-16], the scattering length density of the sample is retrieved. By knowing it's SLD, the unknown layer is detected.

To show the stability of the reconstruction process of the SLD at the presence of the smoothness and roughness of boundaries, the SLD of the sample was retrieved using the data from the variation of surroundings method. As it was described in chapter 3, the method of variation of surroundings would provide us with the information of the mirror image of the original sample with vacuum surroundings at both sides. Using this information as input in Sacks code, consequently would lead to the retrieval of the SLD of the mirror image of the sample.

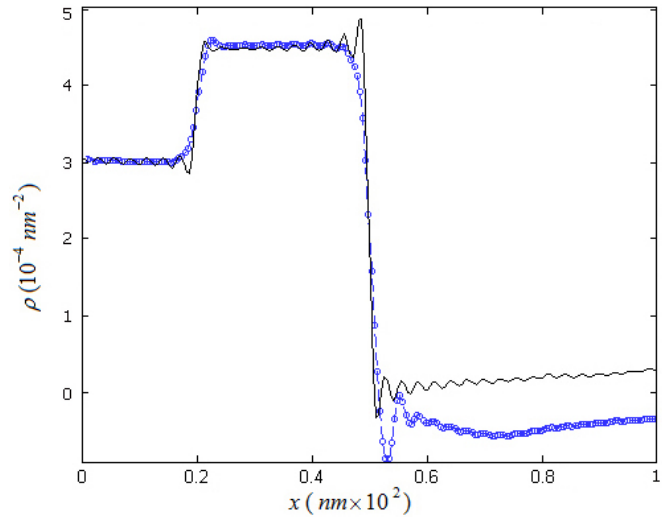

**Figure9:** Reconstructed SLD of the free reversed sample of Fig. 3. Solid line shows the reconstructed SLD of the mirror reversed image of the ideal sample and the circled curve demonstrates the reconstructed SLD of the mirror reversed image of the sample at the presence of smoothness and roughness using variation of surroundings method.

Fig. 9 illustrates the retrieved SLD of the free reversed sample of Fig. 3 by consideration of smoothness and roughness of boundaries. The solid line is the retrieved SLD of the ideal sample while, the circled curve shows the retrieved SLD of the non-ideal sample at the presence of smoothness and roughness of interfaces. The original sample was a 30 nm thick Au with SLD of 4.46 over a 20 nm thick Cr with the SLD of 3.03. However, the mirror image of the sample (20 nm thick Cr over 30 nm thick Au) is clearly demonstrated in Fig. 9. The figure certifies that the method is stable at the presence of smoothness and roughness of interfaces and the output reflectivity data of variation of substrate method for non-ideal samples are still usable to retrieve the SLD of the sample.

## 7 Conclusion

The effects of smoothness and roughness of interfacial potential on the phase of reflection and retrieval process of the SLD of thin films (non-ideal samples) were investigated by using reference method and variation of surroundings with polarized neutrons. In order to investigate the smearing of the optical potentials of two distinct layers at their interfaces, we considered a continuous potential (Eckart function) at the interfaces. The results show that at the presence of smoothness, the output data (reflectivity, phase and polarization of reflected neutrons) for small wave numbers are clearly correspond with the data of the ideal sample, while at large values of *q*, the data deflects from the ideal curve. However in the reference method, some noises were existed in the reflectivity and phase curves which were due to the abrupt change in the elements of *T* matrix (Eq. 22). Thus depending on the range of the neutron wave numbers, consideration of the smoothness is of great importance.

The effects of the roughness of interfaces at the same time with the smoothness were also investigated by random variation of the mean position of the interface by approximately ± 0.5 nm variations around the mean position. To change the position of the interface, a set of random values with normal distribution which were generated by using the Box Muller method, were added (subtracted) to (from) the mean position of the interface and the measurement of reflectivity and phase were performed for each random position. The results show that the effects of roughness are more considerable at large wave numbers. The same as the previous results, several severe noises were existed on the reflectivity curve of the reference method. However, the method of variation of surroundings was free of these abrupt noises at small wave numbers and is more reliable at the presence of roughness.

By using the reflection coefficient data at the presence of smoothness and roughness of interfaces, the SLD of the sample was also retrieved. In order to retrieve the SLD, the data of the reflection coefficient for the whole range of small wave numbers up to *q=0.6* are sufficient and we can ignore the data of large wave numbers. Unfortunately, considering a non-ideal sample with smooth and rough potential at interfaces did not worked well with the reference method because the noises appeared even in the range of small wave numbers, (*q<0.6*). Thus the data of reference method were useless to retrieve the SLD of the sample. However, the results of the variation of surroundings method showed that the method is truly applicable for non-ideal samples and the probable noises would just appear in large wave numbers (*q>0.6*).

In summary, the method of variation of surroundings has the most stability at the presence of smoothness and roughness of interfaces and the retrieved SLD truly corresponds with the ideal sample.